\documentclass[12pt,twoside]{article}   
\usepackage[super,sort,comma]{natbib}
\usepackage{amsfonts}
\usepackage{amsmath}
\usepackage{amssymb}
\usepackage{mathrsfs}
\usepackage{amsbsy}
\usepackage{tensor}
\usepackage{esint}
\usepackage{tikz} 
\usepackage{multirow}
\usepackage{tabularx}
\usepackage{adjustbox}
\usepackage{makecell}
\usepackage{textcomp}
\usepackage{theorem}
\usepackage{epstopdf}
\usepackage[normalem]{ulem}
\usepackage{booktabs,floatrow}

\usepackage{fancyhdr}		

\usepackage[section]{placeins}   %

\usepackage{graphicx}

\makeatletter \renewcommand\@biblabel[1]{$^{#1}$} \makeatother
\newcommand{\cen}[1]{\begin{center} #1 \end{center}}

\usepackage[mathlines]{lineno}

\usepackage{hyperref}
\hypersetup{ colorlinks,
    citecolor=blue,
    filecolor=blue,
    linkcolor=blue,
    urlcolor=blue
}

\usepackage{xcolor}

\definecolor{gray}{rgb}{0.6,0.6,0.6}
\definecolor{red}{rgb}{0.85,0,0}
\definecolor{green}{rgb}{0,0.85,0}
\definecolor{blue}{rgb}{0,0,0.85}
\definecolor{beige}{rgb}{0.92,0.87,0.78}
\usepackage[all]{hypcap}    

                    {$\blacksquare$\vspace*{7pt}} 

\def\x{\boldsymbol{x}}
\def\y{{\boldsymbol y}}

\def\v{{\boldsymbol v}}

\def\q{{\boldsymbol q}}
\def\0{{\mathbf{0}}}
\def\bth{{\boldsymbol \theta}}
\def\bphi{{\boldsymbol \phi}}

\def\bdelta{{\boldsymbol \delta}}

\def\R{{\mathbb R}}

\def\Om{\Omega}

\newcolumntype{C}{>{\centering\arraybackslash}p{7.5em}}

\begin{document}

\cen{\sf {\Large {\bfseries Adaptive Multi-resolution Hash-Encoding Framework for INR-based Dental CBCT Reconstruction with Truncated FOV} \\
\vspace*{10mm}
Hyoung Suk Park$^1$ and Kiwan Jeon$^{1,*}$} \\
$^1$National Institute for Mathematical Sciences, Daejeon, 34047, Republic of Korea\\
}

\pagenumbering{roman}
\setcounter{page}{1}
\pagestyle{plain}
Correspondence: Kiwan Jeon Email: jeonkiwan@nims.re.kr \\

\begin{abstract}
Implicit neural representation (INR), particularly in combination with hash encoding, has recently emerged as a promising approach for computed tomography (CT) image reconstruction. However, directly applying INR techniques to 3D dental cone-beam CT (CBCT) with a truncated field of view (FOV) is challenging. During the training process, if the FOV does not fully encompass the patient's head, a discrepancy arises between the measured projections and the forward projections computed within the truncated domain. This mismatch leads the network to estimate attenuation values inaccurately, producing severe artifacts in the reconstructed images. In this study, we propose a computationally efficient INR-based reconstruction framework that leverages multi-resolution hash encoding for 3D dental CBCT with a truncated FOV. To mitigate truncation artifacts, we train the network over an expanded reconstruction domain that fully encompasses the patient’s head. For computational efficiency, we adopt an adaptive training strategy that uses a multi-resolution grid: finer resolution levels and denser sampling inside the truncated FOV, and coarser resolution levels with sparser sampling outside. To maintain consistent input dimensionality of the network across spatially varying resolutions, we introduce an adaptive hash encoder that selectively activates the lower-level features of the hash hierarchy for points outside the truncated FOV. The proposed method with an extended FOV effectively mitigates truncation artifacts. Compared with a naive domain extension using fixed resolution levels and a fixed sampling rate, the adaptive strategy reduces computational time by 60\% for an image volume of $800\times800\times600$, while preserving the PSNR within the truncated FOV.
\end{abstract}

\newpage     


\newpage

\setlength{\baselineskip}{0.7cm}      

\pagenumbering{arabic}
\setcounter{page}{1}
\pagestyle{fancy}
\section{Introduction}
Dental cone-beam computed tomography (CBCT) has gained popularity as a cost-effective, low-radiation alternative to multi-detector computed tomography (MDCT) in dental clinics \cite{SCARFE2008,LUDLOW2008,KAASALAINEN2021}. It is widely used in various dental applications, including implant planning, orthodontic assessment, and orthognathic surgical planning \cite{Elnagar2020}. Typically, dental CBCT systems use small flat-panel detectors with slower scanning speeds, enabling lower cost and compact system designs that require significantly less space than MDCT systems. However, this design often results in a truncated field of view (FOV) that does not fully encompass the patient's head, leading to missing projection data and making the corresponding inverse problem ill-posed \cite{Park2024}.

The Feldkamp-Davis-Kress (FDK) algorithm \cite{FELDKAMP1984} has been widely used for 3D CBCT reconstruction due to its simplicity, computational efficiency, and ease of implementation. It operates by first applying one-dimensional filtering along the horizontal axis of the flat-panel detector to the measured projection data, followed by a weighted back-projection. However, the FDK algorithm exhibits significant limitations in handling incomplete or missing projection data, often resulting in severe data-truncation artifacts and degraded image quality \cite{Park2024}.

Recently, implicit neural representation (INR) has emerged as a promising alternative for CT image reconstruction, representing images as continuous functions parameterized by neural networks. INR-based reconstruction effectively captures complex spatial relationships between image pixels or voxels, significantly reduces the dimensionality of the solution space, and thereby enables more accurate and efficient reconstruction from incomplete projection data, such as sparse views \cite{Kim2022,Shen2022,ZHA2022} or truncated projections \cite{Park2025}. To facilitate the learning of high-frequency details in CT images, most studies have employed either positional encoding \cite{Tancik2020}, which maps spatial coordinates into a higher-dimensional space using periodic functions, or periodic activation functions \cite{Sitzmann2020}, instead of conventional activations such as ReLU or tanh. In particular, multi-resolution hash encoding \cite{MULLER2022}, popularized by the Instant-NGP framework, has gained significant attention in 3D vision tasks. Hash encoding discretizes the input coordinate space at multiple resolutions and maps it into learnable feature embeddings using a hash table. This approach has demonstrated promising performance in 3D CBCT image reconstruction from sparse-view projection data \cite{Cai2024,Shin2025,ZHA2022}.

However, directly applying INR techniques to 3D dental CBCT with a truncated FOV is challenging. When the FOV does not fully encompass the patient's head, a discrepancy arises between the actual projections and the computed forward projections. Consequently, this mismatch leads to an inaccurate estimation of attenuation values by the network, producing severe artifacts in the reconstructed images (See Figures \ref{fbh_results_comparison} and \ref{hdx_results_comparison}).

In this study, we propose an INR-based approach that mitigates the FOV truncation artifacts by expanding the reconstruction domain to fully encompass the patient’s head. In multi-resolution grid approaches based on hash encoding, naively enlarging the domain using a fixed multi-resolution grid with a large number of resolution levels (hereafter referred to as a fine multi-grid) and applying dense sampling across the entire domain leads to a substantial increase in the computational cost of network training.

To address the computational issue, we adopt an adaptive training strategy that employs a fine multi-grid and dense sampling within the truncated FOV, and coarser multi-grid with sparse sampling outside. Reducing the number of resolution levels decreases the number of trainable hash-encoding parameters, thereby reducing the overall training time. One of our contributions is the introduction of an adaptive hash encoder that selectively activates the lower-resolution features of the hash hierarchy for points outside the truncated FOV. This design maintains a consistent input dimensionality across spatial locations, enabling efficient training with a single unified network architecture without additional modifications.

We evaluate the performance of the proposed method using both numerical phantom and real patient scan. Specifically, we investigate how the resolution levels of the multi-grid and the sampling rate impact the computational cost and the image quality within the truncated FOV.

\section{Method}

\begin{figure*}[!ht]
\centering
\includegraphics[width=1.0\textwidth]{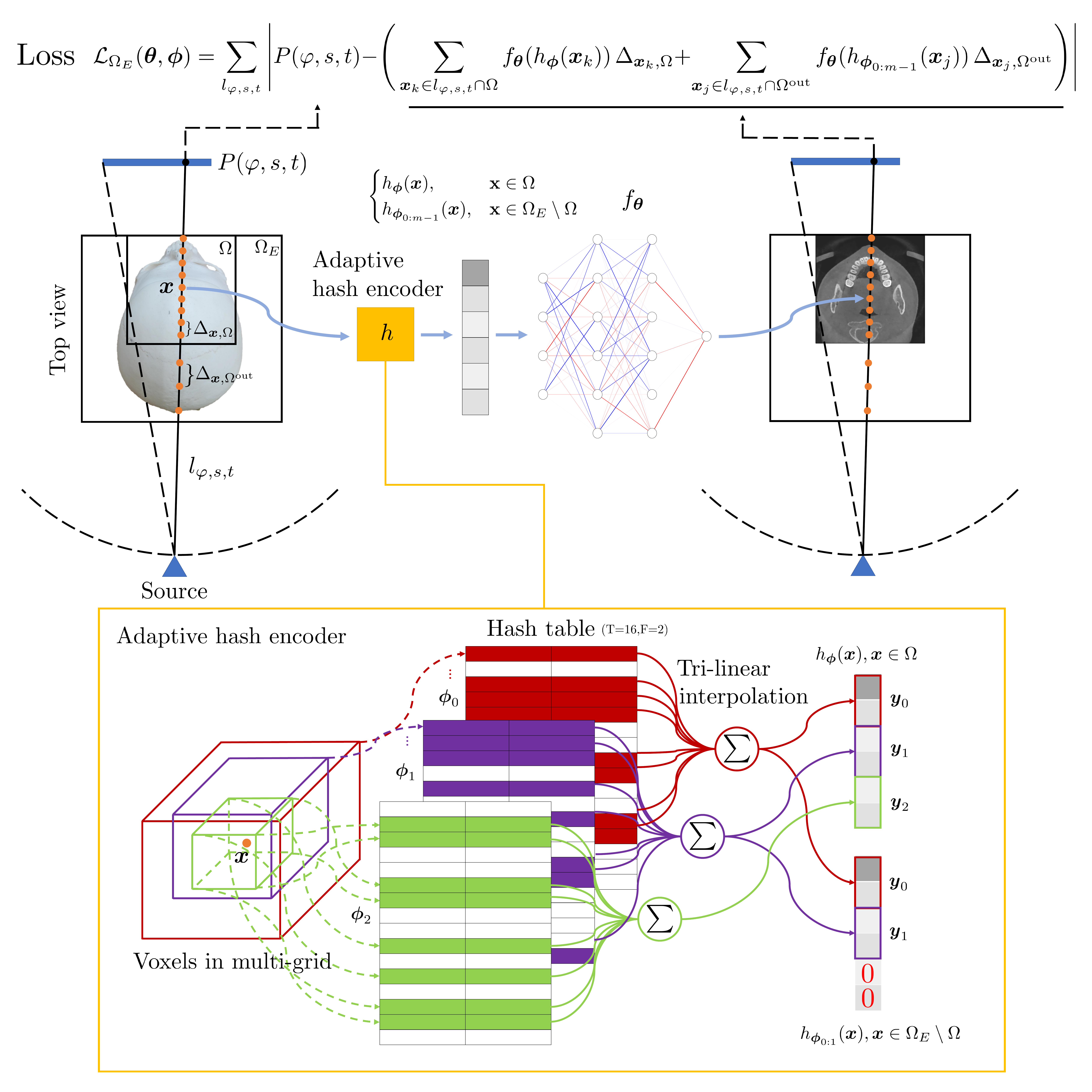}
\caption{Schematic diagram of the proposed INR-based reconstruction framework that leverages adaptive hash encoding for 3D dental CBCT with a truncated FOV ($\Om$). The yellow box illustrates an example of the adaptive hash encoding process with the  number of resolution levels $L = 3$ in $\Om$, the number of resolution levels $m = 2$ in $\Om_E\setminus\Om$, the hash table size $T = 16$, and the feature dimension $F = 2$.}
\label{main_figure}
\end{figure*}

\subsection{INR Reconstruction and FOV Truncation Artifact}
Consider a 3D dental CBCT system equipped with a flat-panel detector. Let $\mu(\x)$ denote the X-ray attenuation coefficient at a spatial position $\x \in \R^3$. According to the Beer-Lambert law~\cite{Beer1852,Lambert1892}, the cone-beam projection data $P(\varphi, s, t)$ measured at detector coordinates $(s,t) \in \R^2$ and projection angle $\varphi \in [0,2\pi)$ can be expressed as
\begin{equation}\label{proj}
P(\varphi, s, t) = \int_{l_{\varphi,s,t}} \mu(\x) dl_{\x},
\end{equation}
where $l_{\varphi,s,t}$ denotes the cone-beam ray connecting the source position at angle $\varphi$ to the detector position $(s,t)$, and $dl_{\x}$ is the line element along the ray.

Let the domain $\Om \subset \R^3$ denote the FOV in the dental CBCT. Given a 3D position $\x\in\Om$, INR learns a neural network $f_{\bth}:\x\mapsto\mu$, parameterized by learnable weights $\bth$. Before directly feeding $\x$ into the network, we apply a multi-resolution hash encoding~\cite{MULLER2022}, denoted by $h_{\bphi}$, which maps $\x$ to hash-encoding parameters $\bphi$ (i.e., a set of learnable features via hierarchical hash tables). Typically, the parameters $\bth$ and $\bphi$ are trained over $\Om$ by minimizing the following objective function:
\begin{equation}\label{inr_loss}
   {\mathcal L_{\Om}(\bth,\bphi)} = \sum_{l_{\varphi,s,t}} \left|P(\varphi, s,t)- \sum_{{\mathbf x}_k\in l_{\varphi,s,t}\cap\Om}  f_{\bth}(h_{\bphi}(\x_k))\,\Delta_{{\mathbf x}_k} \right|,
\end{equation}
where ${\mathbf{x}}_k$ denotes sampled positions along the ray $l_{\varphi,s,t}$ within $\Om$, $\Delta_{{\mathbf{x}}_k} = \|\mathbf{x}_{k+1}-\mathbf{x}_k\|$ is the distance between adjacent sample positions.  The detailed geometry of the 3D dental CBCT system is illustrated in Figure \ref{main_figure}.

However, directly applying these techniques to 3D dental CBCT with a truncated FOV is challenging. When the FOV is truncated and does not fully encompass the patient’s head, there exists points $\x_0\in\Om^C:=\R^3\setminus\Om$ with $\mu(\x_0) > 0$. For any cone-beam ray $l_{\varphi,s,t}$ passing through $\x_0$, we have $\int_{l_{\varphi,s,t} \cap \Omega^C} \mu(\x) \, dl_{\x} > 0$, which implies that 
\begin{equation}
    P(\varphi,s,t) > \int_{l_{\varphi,s,t} \cap \Omega} \mu(\x) \, dl_{\x}.
\end{equation}

During the training process defined by \eqref{inr_loss}, the network matches the entire measured projection $P(\varphi,s,t)$, even though its forward projection only integrates $\mu(\x)$ over the truncated domain $\Om$. This mismatch leads to inaccurate estimation of $\mu(\x)$ within $\Om$, manifesting as severe shadowing artifacts in the reconstructed image, as shown in Figures \ref{fbh_results_comparison} and \ref{hdx_results_comparison}. 

\subsection{INR-based Reconstruction Using Adaptive Hash Encoding}

To address the discrepancy caused by the truncated FOV, we introduce an extended domain, denoted by $\Om_E$, which fully encompasses the patient’s head. However, simply increasing the domain with a fixed high resolution increases the computational time of training the network. 

In this work, we aim to reconstruct morphological details within the domain $\Om$ while accounting for attenuation contributions from the outer domain $\Om^{\text{out}} := \Om_E \setminus \Om$ in a computationally efficient manner. To achieve this, we train the network using a finer multi-grid with dense sampling in $\Om$, and a coarser multi-grid with sparse sampling in $\Om^{\text{out}}$. The use of coarser multi-grid reduces the number of trainable hash-encoding parameters, thereby improving computational efficiency. We construct the adaptive hash encoder to maintain a consistent input dimensionality of the network across spatial locations. Moreover, the use of sparser sampling in $\Om^{\text{out}}$ significantly decreases training time while maintaining image quality within $\Om$.

Specifically, we jointly optimize the parameters $\bth$ and $\bphi$ over the extended domain $\Om_E$ by minimizing the following loss:
\begin{equation}\label{multi_inr_loss}
\begin{aligned}
   &\mathcal{L}_{\Om_E}(\bth,\bphi) 
   =\\ &\sum_{l_{\varphi,s,t}} \Bigg| P(\varphi, s,t) - \Bigg( \sum_{\x_k\in l_{\varphi,s,t}\cap\Om}  
   f_{\bth}(h_{\bphi}(\x_k))\,\Delta_{\x_k,\Om} + \sum_{\x_j\in l_{\varphi,s,t}\cap\Om^{\text{out}}}  
   f_{\bth}(h_{\bphi_{0:m-1}}(\x_j))\,\Delta_{\x_j,\Om^{\text{out}}} \Bigg) \Bigg|,
\end{aligned}
\end{equation}
where the spatial step sizes satisfy $\Delta_{\x,\Om^{\text{out}}} \gg \Delta_{\x,\Om}$, reflecting the sparser sampling in $\Om^{\text{out}}$, and $h_{\bphi_{0:m-1}}$ is a restricted encoder using only the first $m$ levels of the hash hierarchy. In the remaining subsection, we detail the implementation of the adaptive hash-encoding process.

\paragraph{The Proposed Adaptive Hash-Encoding Process}
We now provide a detailed description of the adaptive hash-encoding method applied over the extended reconstruction domain $\Om_E$. The process begins by normalizing $\Om_E$ to the unit cube $[0,1]^3$, and construct $L$ grids at different resolution levels over $\Om_E$. Following the original paper \cite{MULLER2022}, in each level $\ell \in \{0, \dots, L-1\}$, we assign a uniform grid with resolution defined as
\begin{equation}
    N_\ell = \left\lfloor N_{\min} \cdot b^\ell \right\rfloor, \quad \text{where} \quad b = \exp\left( \frac{\ln N_{\max} - \ln N_{\min}}{L - 1} \right),
\end{equation}
where $\lfloor\cdot\rfloor$ denotes floor operation and $N_{\min}$ and $N_{\max}$ denote the coarsest and finest resolutions, respectively. Given an input $\x\in\Om_E$, we scale it by the level-dependent resolution $N_\ell$, as $\x_\ell = \x \cdot N_\ell$.  Then eight integer vertices set of the voxel surrounding $\x_\ell$ in $\mathbb{Z}^3$, denoted as $\mathcal{V}_{\x_{\ell}}$, are spanned as $\mathcal{V}_{\x_{\ell}} = \{ \lfloor \x_{\ell} \rfloor + \bdelta \mid \bdelta \in \{0,1\}^3 \}$. Now, let $\bphi_{l}\in\R^{T\times F}$ be a hash table at level $\ell$, consisting of $T$ learnable  feature vectors of dimension $F$. Each integer vertex $\v_{\x_\ell}=(v_{\x_{\ell,1}},v_{\x_{\ell,2}},v_{\x_{\ell,3}})\in\mathcal{V}_{\x_{\ell}}$ is mapped to an index in the $\bphi_l$ via a spatial hash function $\eta : \mathbb{Z}^3 \to {0, \dots, T - 1}$, given by
\begin{equation}
    \eta(\v_{\x_\ell}) = \left( \bigoplus_{i=1}^3 c_i v_{\x_{\ell,i}} \right) \mod T,
\end{equation}
where $\bigoplus$ denotes the bitwise XOR operation and $c_1=1,c_2=19349663,c_3=83492791$, with $c_2$ and $c_3$ being large prime numbers selected from \cite{MULLER2022}.
For each vertex $\v_{\x_\ell} \in \mathcal{V}_{\x_\ell}$, the associated $F$-dimensional feature vector $\q_{\v_{\x_\ell}}$ is retrieved as $\q_{\v_{\x_\ell}} = \bphi_{\ell}\left[\,\eta(\v_{\x_\ell})\,\right]$. Here, we denote $\bphi_{\ell}[i] \in \mathbb{R}^F$ as the $i$-th row of the matrix $\bphi_{\ell} \in \mathbb{R}^{T \times F}$. Now, the resulting feature vectors corresponding to the eight integer vertices in $\mathcal{V}_{\x_{\ell}}$ are then tri-linearly interpolated by 
\begin{equation}
\y_{\ell} = \sum_{\v_{\x_\ell} \in \mathcal{V}_{\x_\ell}} w_{\v_{\x_\ell}} \q_{\v_{\x_\ell}},
\end{equation}
where $w_{\v_{\x_\ell}}$ is the interpolation weight. For a detailed formulation of the trilinear interpolation weights, we refer to \cite{Levoy1990}. 

Based on the set of $L$-level hash tables $\bphi := \{\bphi_{0}, \dots, \bphi_{L-1}\}$, we construct the adaptive hash encoder as follows. For points $\x \in \Om$, the full multi-resolution encoder $h_{\bphi}$ is constructed by concatenating the interpolated vectors $\y_{0}, \y_{1}, \ldots, \y_{L-1}$ from all levels:  
\begin{equation} 
h_{\bphi}(\x) = [\y_{0}; \y_{1}; \dots; \y_{L-1}] \in \mathbb{R}^{L \cdot F}
\end{equation}
For points $\x \in \Om^{\text{out}}$, we only use the first $m$ levels of $\bphi_{0}, \dots, \bphi_{m-1},~m<L$, and construct a restricted encoder as
\begin{equation}
    h_{\bphi_{0:m-1}}(\x) = [\y_{0}; \dots; \y_{m-1}; \underbrace{\0; \dots; \0}_{L - m \text{ zero vectors}}] \in \mathbb{R}^{L \cdot F},
\end{equation}
where $L-m$ feature vectors $\y_{m}, \ldots, \y_{L-1}$ are replaced with zero vectors to maintain consistent input dimensionality of the network $f_{\bth}$.

\subsection{Data Preparation and Implementation Details}
We evaluated the performance of the proposed method using the FORBILD head phantom~\cite{FORBILD} and a clinical patient scan. In our simulation, the FORBILD head phantom was forward projected using the model described in \eqref{proj}, under the following acquisition settings: a source-to-detector distance of $600.0~\text{mm}$, a source-to-isocenter distance of $400.0~\text{mm}$, a detector resolution of $640 \times 640$ pixels with a pixel spacing of $0.2 \times 0.2~\text{mm}^2$, and $300$ projection views ($n_{\text{views}}$) uniformly distributed over the angular range $[0, 2\pi)$. The current forward model does not account for other physical effects such as beam hardening, scattering, or noise. The FOV $(\Om)$ and the extended FOV $(\Om_E)$ were set to $160.0 \times 160.0 \times 120.0~\text{mm}^3$ and $280.0 \times 280.0 \times 145.0~\text{mm}^3$, respectively. The phantom was entirely enclosed within the extended FOV. 

For the hash encoder, we used parameters similar to those in \cite{MULLER2022}: a minimum resolution $(N_{\min})$ of $16$, a maximum resolution $(N_{\max})$ of $1400$, number of resolution levels $(L)$ of $16$, number of feature vectors $(T)$ of $2^{19}$, and a feature dimension $(F)$ of $2$. In the outer domain ($\Om^{\text{out}}$), we constructed a coarser multi-grid using the first $4$ resolution levels out of $16$. 

The INR network $f_{\bth}$ consists of three fully connected hidden layers with 256 nodes each, whereas the input and output layers consist of 32 (i.e., $L\times F$) nodes and one node, respectively. To ensure that the network outputs non-negative attenuation coefficients, a sigmoid activation function is applied at the output layer. During training, sample points were uniformly drawn along each X-ray with a sampling distance of $0.2$ mm in $\Om$ and $2.0$ mm in $\Om^{\text{out}}$, respectively. The network parameters were optimized using the Adam optimizer with a learning rate of $2\times10^{-4}$ and a batch size of 128 rays. The implementation was based on the PyTorch framework~\cite{PYTORCH}, and training was performed on a system equipped with a CPU (Intel Xeon Gold 6226R 2.90~GHz) and a GPU (Nvidia RTX 3090 24~GB). A CBCT volume of size $800 \times 800 \times 600$ with voxel spacing of $0.2 \times 0.2 \times 0.2~\text{mm}^3$ was reconstructed within $\Om$ using the trained network.

For the clinical experiment, we acquired a dental CBCT scan using a commercial scanner (Q-FACE, HDXWILL, South Korea) with a tube voltage of $85~\text{kVp}$ and a tube current of $8~\text{mA}$. From the measured projection data consisting of 1200 views, 300 projection views were uniformly sampled and used for training. The remaining acquisition settings were similar to those used in the numerical simulation. We also applied the same FOV and extended FOV configurations, as well as the network architecture and hash encoder parameters used in the simulation study.

\section{Results}
\begin{figure*}[!ht]
\centering
\includegraphics[width=1.0\textwidth]{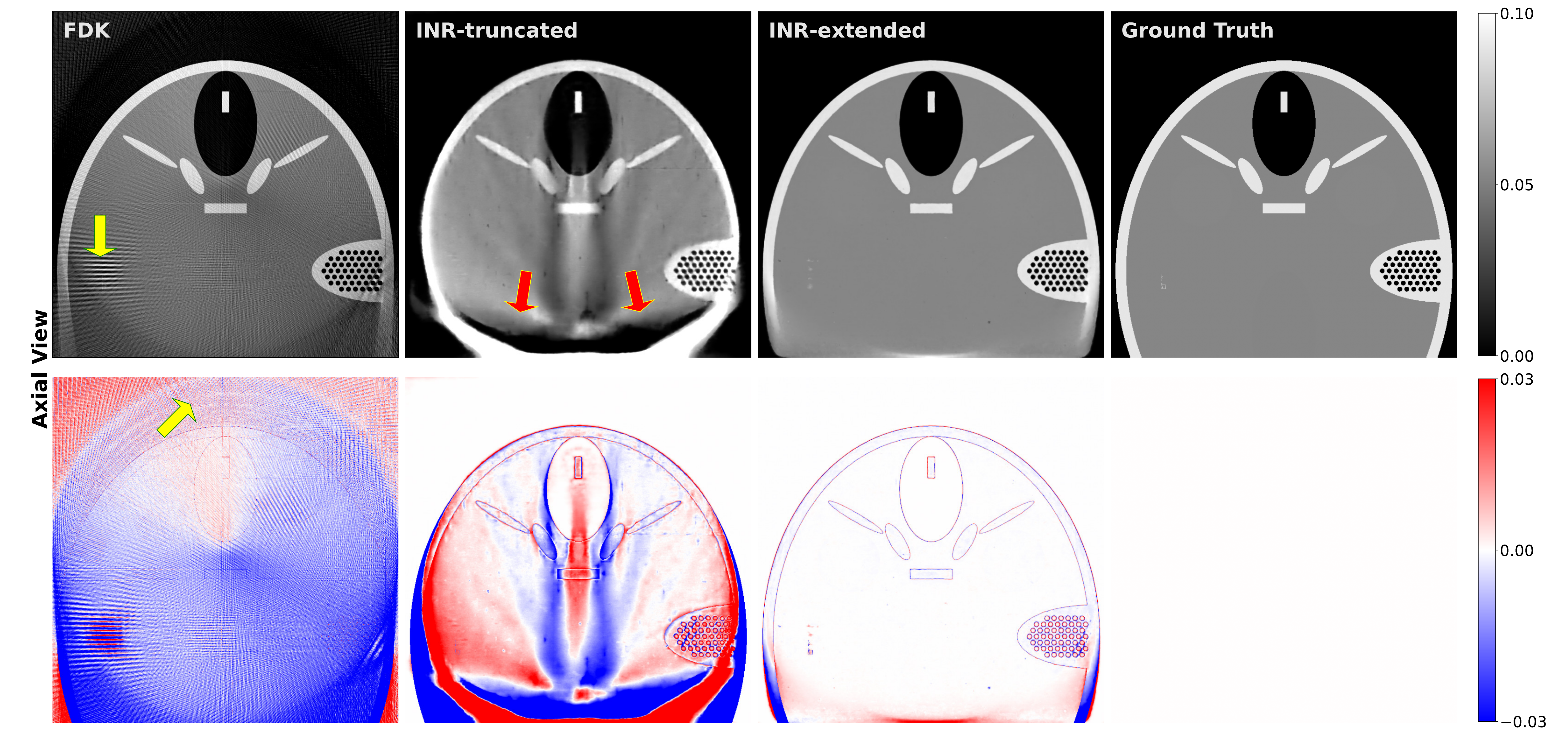}
\includegraphics[width=1.0\textwidth]{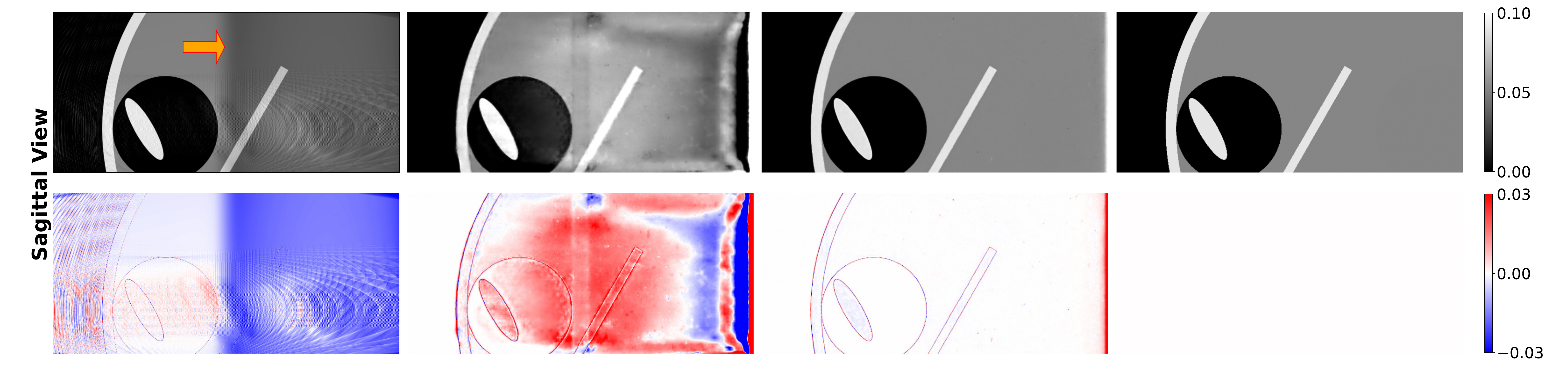}
\caption{Comparison of the reconstruction results for the FORBILD head phantom. The second and fourth rows show the difference images between the reconstructed images and the ground-truth images.}
\label{fbh_results_comparison}
\end{figure*}

\paragraph{Numerical Simulation}
We compared the proposed method with the conventional FDK method and an INR method trained on the truncated FOV (hereafter referred to as INR-truncated) for the FORBILD head phantom. The reconstruction results are presented in Figure~\ref{fbh_results_comparison}. The FDK method does not introduce artifacts caused by forward projections on the truncated reconstruction domain; however, it exhibited noticeable streaking artifacts due to sparse projection views (see yellow arrows). In addition, to mitigate data truncation artifacts in the FDK image, we applied a sinogram extrapolation technique~\cite{Ohnesorge2000}; however, inaccurate extrapolation introduced pronounced negative bias (see orange arrow). The INR-truncated method effectively reduced the streaking artifacts, without introducing bias. However, the INR-truncated method exhibited severe bright and dark shadowing artifacts caused by the truncated FOV (see red arrows). In contrast, the proposed method with the extended FOV effectively suppressed the FOV truncation artifacts. \\

For quantitative evaluation, we computed the peak signal-to-noise ratio (PSNR) and structural similarity index measure (SSIM) with respect to a ground-truth image over the truncated FOV. The quantitative results are illustrated in Table~\ref{tab_quantity}. Among all methods, the proposed approach achieved the highest PSNR and SSIM values. Compared to the INR-truncated method, the proposed method improved the PSNR and SSIM by 14.66 dB and 0.23, respectively.

\begin{table*}[ht]
\caption{Quantitative results of reconstruction methods for the FORBILD head phantom}
\centering
\begin{tabular}{CCCC} 
\toprule %
		&	FDK 		&	INR-truncated 	& 	INR-extended (Proposed) 	\\ \midrule
PSNR	&	39.92 	&	38.98 			&	53.64 			\\ 
SSIM 	&	0.52 	&	0.74 			&	0.97 			\\ 
\bottomrule
\end{tabular}
\label{tab_quantity}
\end{table*}

\begin{figure*}[t]
\centering
\includegraphics[width=1.0\textwidth]{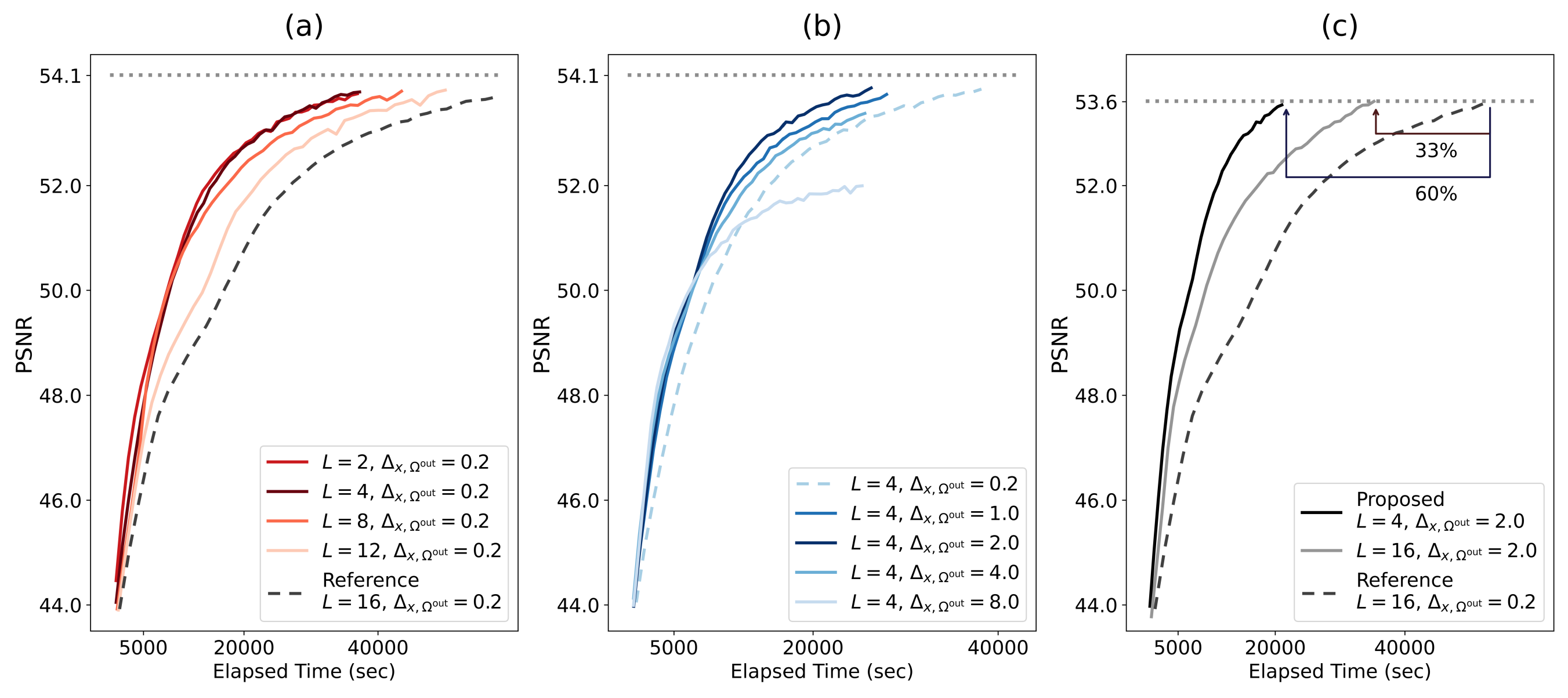}
\caption{Performance comparison of the proposed method for different (a) numbers of resolution levels $L$ and (b) sampling distances $\Delta_{\x,\Omega^{\text{out}}}$ applied in the outer domain $\Omega^{\text{out}}$. The horizontal dotted line denotes the reference PSNR value obtained from a naive domain extension using fixed parameter settings (i.e., $L = 16$, $\Delta_{\x,\Omega^{\text{out}}} = 0.2$). (c) Impact of the resolution level and the sampling distance on training time.  Here, the horizontal dotted line represents the final PSNR value achieved under the adaptive setting ($L = 4$, $\Delta_{\x,\Omega^{\text{out}}} = 2.0$).}
\label{abl_study}
\end{figure*}

Figure~\ref{abl_study}(a) presents the performance of the proposed multi-grid approach with varying number of resolution levels ($L = 2, 4, 8, 12$) applied in the outer domain $\Omega^{\text{out}}$. In this experiment, all networks with different settings were trained for the same number of iterations. As shown in Figure \ref{abl_study}(a), PSNR remained relatively stable across different values of $L$. Among the tested settings, $L = 4$ achieved convergence to a PSNR value close to the reference in the shortest training time. Here, the reference PSNR value is defined as the value obtained from a naive domain extension using a fixed large number of resolution levels and dense sampling (i.e., $L = 16$, $\Delta_{\x,\Omega^{\text{out}}} = 0.2$).

Figure~\ref{abl_study}(b) shows the performance of the proposed method for different sampling distances ($\Delta_{\x,\Omega^{\text{out}}} = 1.0, 2.0, 4.0, 8.0$) applied in the outer domain. In this experiment, the number of resolution levels was fixed at $L = 4$, which was identified as the optimal setting in Figure~\ref{abl_study}(a). As shown in Figure~\ref{abl_study}(b), the PSNR was robust with respect to changes in $\Delta_{\x,\Omega^{\text{out}}}$. The setting $\Delta_{\x,\Omega^{\text{out}}} = 2.0$ achieved convergence to a PSNR value close to the reference in the shortest training time.

Figure~\ref{abl_study}(c) illustrates the impact of the resolution level and the sampling distance on training time. Compared to the naive domain extension using a fixed large number of resolution levels and dense sampling, adjusting both parameters together in the outer domain reduced the training time by 60\%, while resulting in only a slight degradation in the PSNR by 0.92\% (from 54.1~dB to 53.6~dB). In this study, training for the adaptive setting (i.e., $L = 4$, $\Delta_{\x,\Omega^{\text{out}}} = 2.0$) was stopped when the difference in PSNR values computed every $10^4$ iterations fell below $10^{-1}$, yielding a PSNR value of 53.6~dB. For the comparison settings, training was continued until the same PSNR level of 53.6~dB was reached. \\

\paragraph{Clinical Experiment}

\begin{figure*}[!ht]
\centering
\includegraphics[width=1.0\textwidth]{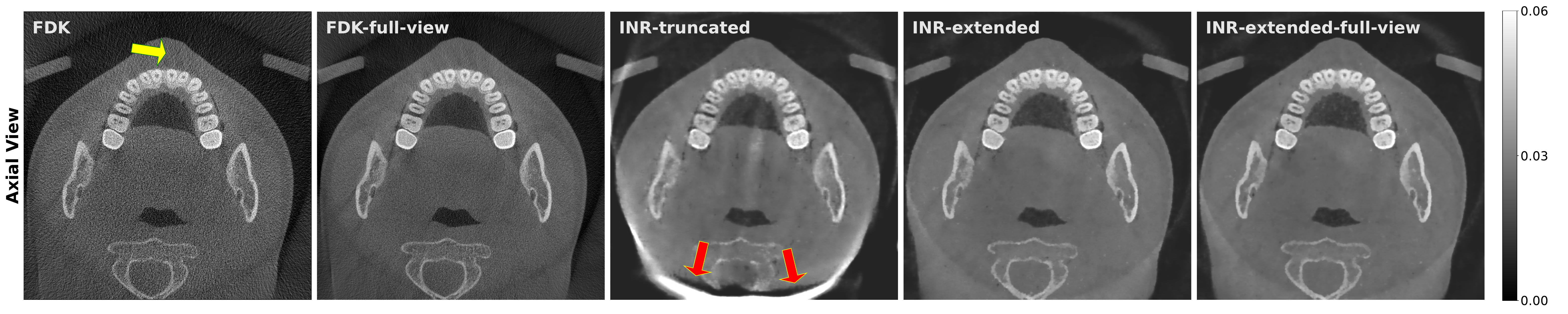}
\includegraphics[width=1.0\textwidth]{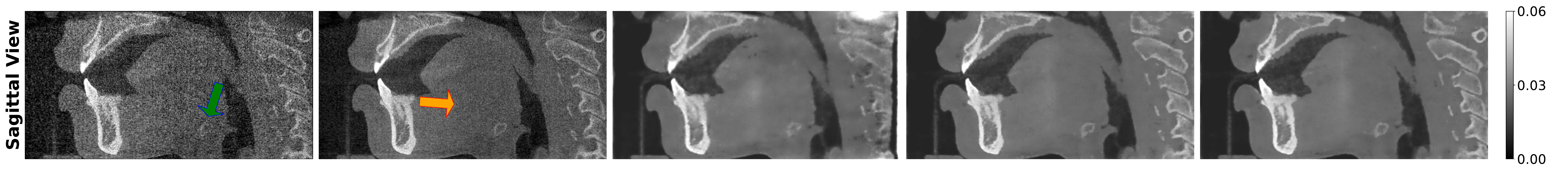}
\caption{Comparison of the reconstruction results for a clinical CBCT scan. The second and fourth rows show the difference images between the reconstructed images and the ground-truth images.}
\label{hdx_results_comparison}
\end{figure*}

Figure~\ref{hdx_results_comparison} presents a comparison among the proposed method, the conventional FDK method, and the INR-truncated method for a clinical CBCT scan. Since ground-truth images are not available in the clinical setting, we additionally include reconstructions from both the FDK and the proposed methods using full-view projection data ($n_{\text{views}}=1200$) as reference images.

As shown in Figure~\ref{hdx_results_comparison}, the proposed method effectively suppressed FOV truncation artifacts compared to the INR-truncated method (see red arrows). Relative to the FDK method, the proposed method more effectively reduced the streaking artifacts caused by sparse projection views, while better preserving morphological details (see yellow and green arrows). Although the FDK reconstruction with full-view data reduced streaking artifacts and preserved fine structures, it still exhibited noticeable bias due to the inaccurate sinogram extrapolation technique (see orange arrow). In contrast, the proposed method—both with sparse- and full-view projections—successfully mitigated such bias.\\

\section{Discussion and Conclusion}
In this study, we proposed a computationally efficient INR-based reconstruction framework that leverages multi-resolution hash encoding for 3D dental CBCT with a truncated FOV. To mitigate the truncation artifacts resulting from truncated FOV, we trained the network over an expanded reconstruction domain to fully encompass the patient’s head. To reduce computational burden, we employed a fine multi-grid and dense sampling inside the truncated FOV, and coarser multi-grid with sparse sampling in the outer domain. We introduced an adaptive hash encoder to ensure consistent input dimensionality across spatially varying resolutions, enabling training using a single unified network architecture.

Our numerical experiments using the FORBILD head phantom demonstrated that the proposed method substantially reduced truncation artifacts while preserving fine morphological details within the FOV (see Figure~\ref{fbh_results_comparison}). Compared to naive domain extension using a fixed multi-resolution hierarchy and dense sampling across the entire extended FOV, selectively adjusting both the resolution levels and the sampling parameters in the outer domain led to a significant reduction in training time, without compromising reconstruction quality within the FOV in terms of PSNR (see Figure~\ref{abl_study}). These results demonstrate that the effects of data mismatch due to the truncated FOV can be effectively mitigated through a computationally efficient strategy. The clinical experiments further validated the practical applicability of the proposed method (see Figure \ref{hdx_results_comparison}).

Similar truncation artifacts also arise in iterative reconstruction approaches. During the reconstruction process, if the FOV does not fully encompass the patient’s head, the discrepancy between the measured projections and the forward projections computed over the truncated FOV accumulates over iterations, leading to pronounced artifacts in the reconstructed images. Previously, Dang {\it et al}. \cite{DANG2017} proposed a multi-resolution iterative reconstruction method with an extended FOV to mitigate the truncation artifacts caused primarily by the head holder in their prototype CBCT system. Although their approach shows effectiveness in certain settings, its applicability to dental CBCT remains limited due to the compact size of flat-panel detectors. In such settings, the extended anatomical region beyond the detector coverage yields severely missing projection data. Consequently, even with conventional regularization techniques such as total variation, the iterative reconstruction produces degraded image quality \cite{Park2025b}. In contrast, the proposed method more effectively addresses the missing projection data by leveraging the superior representational power of INRs.

The conventional FDK reconstruction method is an analytic approach and is therefore inherently free from artifacts caused by forward projections on the truncated reconstruction domain. However, when the projection data are truncated or missing due to the truncated FOV, the analytic FDK method produces severe bright streaking and shadowing artifacts \cite{Ohnesorge2000}. To address these artifacts, various methods have been proposed over the past few decades. These include sinogram extrapolation techniques \cite{Ohnesorge2000,HSIEH2004}, alternative reconstruction algorithms such as differentiated backprojection \cite{Defrise2006,Noo2004,YU2006}, and variational approaches that incorporate regularization priors such as total variation \cite{Yu2009}, generalized L-spline \cite{Lee2015}, and deep learning \cite{Han2022,Ketola2021} 

The proposed mehtod has some limitations. First, we assumed that the extended reconstruction domain is selected to fully cover the patient’s head. In clinical settings, however, this extent is not known a priori and should be carefully determined with consideration for computational constraints. Second, the optimal selection of resolution levels and sampling density may vary depending on the acquisition settings. Future research could explore how variations in the acquisition domain influence these parameters. Finally, while the proposed method improves computational efficiency through adaptive encoding, it still requires considerable computational effort for clinical use. To further reduce training time and computational cost, advanced strategies, such as employing meta-learning frameworks to initialize the MLP network with effective weights \cite{Nichol2018} and incorporating hash-encoding regularization \cite{Shin2025}, could be integrated into the proposed framework as promising future extensions.

\section*{Acknowledgment}
H.S.P. and K. J were supported by  the National Institute for Mathematical Sciences (NIMS) grant funded by the Korean government (No. NIMS-B25910000). H.S.P. and K.J. were partially supported by the National Research Foundation of Korea(NRF) grant funded by the Korea government(MSIT) (No. RS-2024-00338419).

\section*{Conflict of Interest Statement}
The authors declare no conflicts of interest.

\section*{References}

\end{document}